\documentstyle[12pt,aaspp4,flushrt]{article}

\received{}
\accepted{}

\slugcomment{}

\lefthead{Bower et al.}
\righthead{Nucleus of M84}

\begin{document}

\title{Kinematics of the Nuclear Ionized Gas in the
Radio Galaxy M84 (NGC~4374)\altaffilmark{1}}

\author{G. A. Bower\altaffilmark{2}, R. F. Green\altaffilmark{2}, 
A. Danks\altaffilmark{3},
T. Gull\altaffilmark{4}, S. Heap\altaffilmark{4}, J. Hutchings\altaffilmark{5},
C. Joseph\altaffilmark{6}, M. E. 
Kaiser\altaffilmark{4,7}, R. Kimble\altaffilmark{4}, S. Kraemer\altaffilmark{8},
D. Weistrop\altaffilmark{9}, B. Woodgate\altaffilmark{4}, D. 
Lindler\altaffilmark{10}, R. S. Hill\altaffilmark{3}, E. M. Malumuth\altaffilmark{3}, S. Baum\altaffilmark{11},
V. Sarajedini\altaffilmark{12}, 
T. M. Heckman\altaffilmark{7,11},
A. S. Wilson\altaffilmark{11,13}, and D. O. Richstone\altaffilmark{14}}

\vskip 36truept

\centerline{To appear in the Astrophysical Journal {\it Letters}}


\altaffiltext{1}{Based on observations with the NASA/ESA {\it Hubble Space Telescope},
obtained
at the Space Telescope Science Institute, which is operated by the
Association of Universities for Research in Astronomy, Inc. (AURA), under
NASA contract NAS5-26555.}

\altaffiltext{2}{Kitt Peak National Observatory, National Optical Astronomy Observatories, P. O. Box 26732,
Tucson, AZ 85726}
\altaffiltext{3}{Hughes/STX, NASA/Goddard Space Flight Center, Code 681, Greenbelt, MD 20771}
\altaffiltext{4}{NASA/Goddard Space Flight Center, Code 681, Greenbelt, MD 20771}
\altaffiltext{5}{Dominion Astrophysical Observatory, 
National Research Council of Canada,
5071 W. Saanich Road, 
Victoria, BC V8X 4M6, Canada}
\altaffiltext{6}{Dept. of Physics \& Astronomy, Rutgers University,
P. O. Box 849, Piscataway, NJ 08855}
\altaffiltext{7}{Department of Physics \& Astronomy, Johns Hopkins University,
Homewood Campus, Baltimore, MD 21218}
\altaffiltext{8}{Catholic University, Department of Physics, NASA/Goddard Space Flight Center, Code 681, Greenbelt, MD 20771} 
\altaffiltext{9}{Department of Physics, University of Nevada, 4505 S. Maryland
Parkway, Las Vegas, NV 89154}
\altaffiltext{10}{Advanced Computer Concepts, Inc., NASA/Goddard Space Flight Center, Code 681, Greenbelt, MD 20771}
\altaffiltext{11}{Space Telescope Science Institute, 3700 San Martin Drive,
Baltimore, MD 21218}
\altaffiltext{12}{Steward Observatory, University of Arizona, Tucson, AZ
85721}
\altaffiltext{13}{Astronomy Department, University of Maryland, College Park,
MD 20742}
\altaffiltext{14}{Department of Astronomy, University of Michigan, Dennison Building, Ann
Arbor, MI 48109}


\newpage

\begin{abstract}

We present optical long-slit spectroscopy of the nucleus of the nearby
radio galaxy M84 (NGC~4374 = 3C~272.1) obtained with the
Space Telescope Imaging Spectrograph (STIS) 
aboard the {\it Hubble Space Telescope} (HST). 
Our spectra reveal
that the nuclear gas disk seen in WFPC2 imaging
by Bower et al.~(1997, ApJ, 483, L33) is rotating rapidly. The velocity
curve has an S-shape with a peak amplitude of 400 km~s$^{-1}$
at $0\farcs1 = 8$ pc from the nucleus. To model the observed gas kinematics,
we construct a thin Keplerian disk model that fits the data well
if the rotation axis of the gas disk is aligned with the radio jet axis.
These models indicate that the gas dynamics are driven by a nuclear compact
mass of $1.5 \times 10^9 \ M_{\odot}$ with an uncertainty range of 
$(0.9 - 2.6) \times 10^9 \ M_{\odot}$ and that the inclination of
the disk with respect to the plane of the sky is $75\arcdeg - 85\arcdeg$.
Of this nuclear mass, only $\le 2 \times 10^7 \ M_{\odot}$
can possibly be attributed to
luminous mass. Thus, we conclude that a dark compact mass (most
likely a supermassive black hole) resides in
the nucleus of M84.

\end{abstract}




%

\newpage   

\section{Introduction}

M84 is an E1 galaxy in the Virgo Cluster with an active 
galactic nucleus 
and hosts the F-R I (Fanaroff \& Riley 1974) radio source 3C~272.1. 
Bower et al.~(1997; hereafter
Paper~I) obtained 
images of M84 with the Wide
Field/Planetary Camera 2 (WFPC2) aboard HST, showing that the ionized gas
within the central kpc has three components: a nuclear gas disk, outer
filaments, and an
`ionization cone'. The nuclear gas disk has
diameter $\approx 1''$ (82 pc) and a major axis P.A. $\approx 58\arcdeg$
that is tilted by $\approx 25\arcdeg$ with respect to the major axis P.A. of the outer filamentary
emission. This outer filamentary emission had been seen
in ground-based imaging (e.g., Hansen et al.~1985; Baum et al.~1988). Its
major axis is approximately
perpendicular to the axis of the radio jets  
(Laing \& Bridle 1987;
Jones et al.~1981).

The presence of a nuclear gas disk in M84 is especially interesting.
If the gas exhibits Keplerian motion about the nucleus, then a straightfoward
application of Newton's laws to 
the dynamics of this gas disk would provide an estimate of the
mass of the putative supermassive black hole (BH) in M84's nucleus.
It is plausible that M84 contains a BH, since it 
is a radio galaxy and the rotation gradient of the ionized gas is
spatially unresolved (i.e., $> 100$ km~s$^{-1}$ arcsec$^{-1}$) in ground-based observations
(Baum et al.~1990, 1992). Previous HST observations using the
Faint Object Spectrograph have found gas-dynamical evidence for BHs in other galaxies
containing nuclear gas disks, such as M87 and NGC~4261 (Harms et al.~1994;
Ferrarese et al.~1996). STIS (through
the use of
a CCD in a long-slit spectrograph) provides a significant improvement in 
HST's efficiency for measuring the nuclear dynamics of galaxies. We chose
M84 as a target for a demonstration.

In this paper, we analyze the dynamics of the nuclear
gas disk in M84 to probe the nuclear gravitational potential. The
analysis of gas dynamics in galactic nuclei is complementary to the method of
analyzing the stellar dynamics (e.g., Kormendy et al.~1996). However,
measuring the stellar kinematics in M84 using HST would be challenging, since the
stellar surface brightness at the nucleus (determined from the WFPC2 
F547M image in Paper~I) is rather modest ($\mu_V
\approx$ 16 mag arcsec$^{-2}$). For M84, the gas kinematics are far easier to measure
than the stellar kinematics given the high surface brightness in the 
emission lines.   
  
Throughout this paper, we adopt a distance to M84 of 17 Mpc 
(Mould et al.~1995). At this distance, 
$1''$ corresponds to 82 pc. The 
Galactic extinction along the line of sight is
$A_B = 0\fm13$
(Burstein \& Heiles 1984).

\section{Observations and Data Calibration}

Long-slit spectroscopy of M84's nuclear region was obtained with the STIS
CCD, which has a pixel scale of $0\farcs05$/pixel
(Baum et al.~1996),
aboard HST on 1997 April 14 and 17 with the telescope tracking in fine lock
with one FGS probe 
(nominal jitter $\approx 0\farcs007$). Since M84's nucleus contains a
bright
optical point source (Paper~I), the nucleus was acquired easily to
an accuracy of $0\farcs05$ by
the ACQ mode using two iterations of two 10 sec imaging exposures through
the F28X50LP optical long-pass filter. The
ACQ/PEAK mode (Baum et al.~1996) was not available during these observations since
they were obtained early during Servicing Mission Orbital Verification (SMOV). 
The $52'' \times 0\farcs2$ slit was aligned at a position angle (P.A.) of
$104\arcdeg$. This was the closest that the slit could be aligned with the
gas disk's major axis (P.A. $= 58\arcdeg$; Paper~I) because of HST scheduling constraints
during SMOV.
To allow for the centering accuracy of only $0\farcs05$ and for the offset
between the slit P.A. and the gas disk's major axis, we planned to
obtain spectra at
four different slit positions offset from the nucleus by
$-0\farcs3$, $-0\farcs1$, $+0\farcs1$, and $+0\farcs3$, where the 
offsets were perpendicular to the slit and negative
spatial offsets moved the slit toward a P.A. of $14\arcdeg$ on the sky.
However, below we discuss an empirical determination of
the slit positions using our STIS data and the WFPC2 images from Paper~I,
showing that the actual offsets were $-0\farcs2$,
$0\farcs0$, $+0\farcs2$, and $0\farcs0$. For the fourth offset position,
the discrepancy between the planned and actual positions occurred
because this last spectrum
was obtained during the second visit of M84, with an erroneously commanded
offset. This error was fortuitous
because from our analysis below it is apparent that the kinematic signature
of the
nuclear gas disk is readily detectable only within $\sim 0\farcs3$ of the
nucleus, beyond which the kinematics of the outer filamentary emission (which do not
necessarily provide good leverage on the nuclear gravitational potential) dominates the spectrum.
  
At each slit position, we obtained spectra with
the G750M grating, which has a dispersion of 0.56 \AA/pixel.
This grating was set to cover the wavelength range of 6295~\AA \ to
6867~\AA, which includes the emission lines of H$\alpha$, 
[N~II] $\lambda\lambda
6548, 6583$, and [S~II] $\lambda\lambda 6717, 6731$. The spectral resolution
of our instrumental configuration was 2.2 \AA \ $\approx 100$ km~s$^{-1}$
(FWHM), assuming uniform illumination of the slit. However, Paper~I shows
that at the nucleus, there is a point source in the optical continuum, and
the H$\alpha$ + [N~II] emission is very compact. Thus, our spectral resolution at the nucleus was
better than 100 km~s$^{-1}$. We integrated for two HST orbits at each slit
position, which was equivalent to $4500 - 5100$ sec per slit
position depending on the occurrence of instrumental overheads. Spectra of the internal wavelength calibration source (wavecals)
were
interspersed among the galaxy spectra to allow for correction of
thermal drifts during
data reduction.

The data were calibrated using the CALSTIS pipeline to
perform the steps of bias subtraction, dark subtraction, applying the
flatfield, and combining the two sub-exposures to reject cosmic-ray events.
The accuracy of the flatfield calibration was 1\%.
To reject hot pixels from the data, we employed dark frames 
obtained immediately before and after the M84 observations. We examined the
input data, the flagged hot pixels, and the cleaned output data to
ensure that only hot pixels were rejected. The data were
wavelength calibrated and rectified by tracing the wavecals (using the
Ne emission lines for the dispersion axis and the shadows of the two
occulting bars for the spatial axis) and then applying these solutions for 
the geometric distortions to the data. The largest offset that we found between
the actual and nominal dispersion solutions was 0.30 pixels, which is 
significant given that the dispersion solutions were accurate to
0.06 pixels (0.03 \AA \ = 1 km~s$^{-1}$).
The data were then rebinned onto a log $\lambda$ scale with a reciprocal dispersion
of 25 km~s$^{-1}$~pixel$^{-1}$.

At this point, we measured the slit positions implied by comparing
the flux along the slit with the WFPC2 images from Paper~I. Since the
on-orbit flux calibration for our grating tilt was not known accurately
at the time
of our analysis, we normalized the spectra by the observed continuum intensity
distribution.
The predicted continuum normalized WFPC2 F658N fluxes were then determined by using
the STSDAS task ``synphot'' on the normalized STIS spectra. These predictions
were compared with the
WFPC2 F658N image normalized by the synthetic 6590~\AA \ continuum
image (constructed from the F547M and F814W images; see Paper~I for
details), indicating that the actual slit positions were offset by $-0\farcs2$,
$0\farcs0$, $+0\farcs2$, and $0\farcs0$ (shown in Fig.~1). However, the uncertainty is as high as $\pm 0\farcs1$ since
the nuclear gas disk is very compact (Paper~I) and the slit width
was rather large.
These empirical positions are favored because
the
gas kinematics (measured in \S 3) are symmetric with respect to the nucleus
for these empirical positions. The relative steps between slit positions
of $0\farcs2$ are assumed to be much more accurate than the absolute
placement.

\section{Measurement of the Gas Kinematics}

Fig.~2 shows the last offset = $0\farcs0$ spectrum
centered on the [N~II] $\lambda 6583$ emission line.
The first iteration at measuring the radial velocities involved cross 
correlating each spectral row 
from the four long-slit spectra with a synthetic emission-line spectrum, which
included only the
five emission lines that we detected in M84 (see Fig.~3)
with flux ratios set at
values typically found in the data. We then compared the velocities
measured by the cross correlation technique with those measured
manually from the emission-line peaks. These measurements agree very well for distances $> 0\farcs3$ from the nucleus.
For rows closer to the nucleus than this, the emission line profiles 
usually exhibit
two kinematic components rather than a single component. These two components are readily seen in the strong [N~II] $\lambda 6583$ 
profile (as shown in Fig.~4), especially for the two spectra with
offset = $0\farcs0$. Since the velocity measured by our cross correlation technique
(using our synthetic template spectrum) 
coincides
with the flux-weighted centroid over a
given
emission-line profile, these measurements will be distorted when more than
one kinematic component is present.
We determined which of the [N~II] $\lambda 6583$ profiles in Fig.~4 have
two kinematic components by fitting model profiles to a few examples.
Separate models with one or two Gaussians were fit to the profiles.
If the improvement in $\chi^2$ was significant, then the profile was 
classified as having two kinematic components. 
Although these models
were not good fits to the observed profiles (which have broader
wings than a Gaussian), this procedure was sufficient for objectively 
determining which
profiles have two components. The velocities for the separate components
were then measured by finding the centroid of each component peak.

Based on the H$\alpha$ + [N~II] image (see Fig.~1), it is not
surprising that two kinematic components are seen within $\sim 0\farcs3$ of
the nucleus in the STIS spectroscopy. Paper~I identified three spatial
components in the ionized gas, including the nuclear gas disk, the outer filaments, and an ionization cone. Given these spatial components, the line of sight
to the nuclear gas disk should also intersect the outer filamentary gas 
(and perhaps the ionization cone) lying
in the foreground. Since the outer filamentary gas rotates about the
nucleus at $\approx \pm 100$ km~s$^{-1}$ (Baum et al.~1990, 1992), one expects
to see this low-velocity component superposed onto the high-velocity
kinematics of the nuclear gas disk. Fig.~5 shows our velocity
measurements along the slit for each of the four slits. Our measurements
of the low-velocity component agree very well with those of Baum et al.~(1990).
This measurement of the high-velocity component is the first time that the
kinematics of M84's nuclear gas disk have been resolved.

\section{Interpretation}

The high amplitude and S-shape of the velocity curve in the center of
M84 strongly suggest that the emitting gas is in organized motion around a
massive central object.  To estimate the central mass, numerical models were
constructed of a thin Keplerian disk.  The continuous velocity sampling along
the slit and the two offset slit positions give important constraints on the
model parameters.

The models were calculated as follows.  By construction, $V$ (the velocity of gas at
any point on the disk) is proportional to $r^{-1/2}$, where $r$ is the distance
from the center in the disk plane.  The observed velocities were matched by
adopting a systemic velocity for the emitting gas. The relative weighting as
a function of radius was assumed to follow the major axis isophotal intensity map 
$I(r)$ derived
in Paper~I (i.e., $I(r) \propto r^{-1}$).  To
avoid divergence, the velocity was truncated interior to a selected physical radius
at its value at that radius; the intensity weighting was truncated
interior to a chosen projected radius at its corresponding value. The
disk is inclined to the plane of the sky by an angle $i$, and the slit is
rotated with respect to the major axis of the disk by an angle $\theta$.  The
slit can also be offset from the nucleus by a specified distance in a direction
perpendicular to the slit rotation angle. The
contribution to the velocity peak at any sample pixel is weighted by both the
projected disk intensity distribution and by the PSF of HST + STIS.  The PSF was approximated as a Gaussian with dispersion
$\sigma$, centered at the midpoint of each sample pixel and integrated to a distance of $4\sigma$.  Since the $0\farcs2$ slit
is substantially wider than the FWHM of the delivered PSF, the measured
velocity centroid of each pixel sample should be dominated by the effects of the PSF.
 
The most important free parameter is the mass of the central object.  The
range of the other parameters is interesting primarily for the extent to which
they bound the acceptable range of masses, with one exception to be discussed
below.  After many realizations of the model, several of the parameters were
fixed, either because they were tightly or externally constrained, or because
the results were insensitive to reasonable choices.  The physical inner
radius and the intensity weighting inner radius are examples of the latter case.  The PSF is known from both modeling and
measurement of standard stars.  A Gaussian fit to a model encircled
energy curve for HST + STIS + guiding jitter suggested $\sigma = 0\farcs06$.
A fit to a short exposure of a standard star observed with the G750M
grating produced $\sigma = 0\farcs04$.
Those values bracketed the actual width of the velocity inflection of the
S-shaped curve.  A $\sigma = 0\farcs05$ was adopted, corresponding to a FWHM of
$0\farcs12$.  The fact that the PSF for a longer exposure guided with just one
FGS probe appears to be somewhat wider than that for a brief standard star
exposure is not unreasonable.  The systemic velocity required to symmetrize
the peak velocities of the model with respect to the data is 1125 km~s$^{-1}$.  A
variation of 25 km~s$^{-1}$ in either direction leads to a noticeable degradation
in the quality of the fit.
 
The first approach to determining the central mass was to assume that
the isophotal fit of the emission-line intensity in Paper~I gave directly the ellipticity
and P.A. of the gas disk within $0\farcs5$ of the nucleus.  Those
values are an ellipticity of 0.17, corresponding to $i = 34\arcdeg$, and a P.A. of $58\arcdeg$, leading to
a relative slit angle of $46\arcdeg$.  No acceptable fit of the zero-offset
positions could be produced at $i = 34\arcdeg$.  The modeled velocities of the
``Keplerian'' portion of the curve, which are particularly well represented in
the data in the negative direction as plotted in Fig.~5, returned to zero
much more slowly than the data.  An acceptable fit to the zero-offset
positions could be obtained by assuming that the ellipticity of the isophotes
did not represent the inclination angle of the gas disk.  A much
higher inclination near $75\arcdeg$ was required in that case, and no corresponding fit to
the offset positions at $+0\farcs2$ and $-0\farcs2$ could be produced.  The velocity
amplitudes at the point of slit closest approach to the nucleus were very
much higher than observed and the shapes were not at all congruent.
 
This result led us to question whether the abrupt isophote twist in the
inner arcsecond actually represented the major axis of the gas disk.
An alternative approach is to assume that the major axis of the larger scale
emission-line structure defines the P.A. of the kinematic major axis of the
inner gas disk.  Baum et al.~(1988) measured that angle on the larger scale
to be $83\arcdeg$.  That
direction is nearly perpendicular to the measured P.A.'s of the radio
jets of 3C 272.1, which are $-5\arcdeg$ and $+170\arcdeg$ (Laing \& Bridle 1987).
That choice fixes the relative slit angle at $21\arcdeg$.  The free parameters
are then the central mass and inclination angle.  The best fit model is
plotted over the data in Fig.~5, while its parameters (with uncertainty
ranges) are given in Table~1.  Fig.~5 shows reasonable
agreement with the velocity amplitude and Keplerian shape for the zero-offset
curves, and qualitative correspondence with the off-nuclear curves.
Our Keplerian disk model fits the data very well, although there may
be minor non-disk contributors to the emission-line profile (as suggested
by the complexities of the inner emission-line isophotes [Paper~I] and
spectral profiles [Fig.~4]).  

The acceptable range of the model parameters should be determined
quantitatively.  However, a proper derivation of $\chi^2$ per degree of freedom is
complicated by two factors: the data from adjacent pixels are strongly
correlated through the PSF, and the uncertainties in velocity points are dominated by
systematic effects in disentangling the multiple broad components.  In
addition, the derived central mass and disk inclination angle are themselves
correlated.  As the disk becomes more edge-on, the observed radial velocities
more nearly represent the total velocity, implying a smaller mass for a given
amplitude.  That effect is overwhelmed by a competing effect in a thin disk
model, which is the foreshortening in the transverse direction.  A given
pixel samples to much larger radii as the disk becomes edge-on, requiring
larger mass to produce a given velocity amplitude.  Increasing inclination
angle therefore requires increasing mass.  To set bounds on the acceptable
range of mass, we sampled the two corners of the error ellipse.  These
represent approximately factors of 2 increase in $\chi^2$, but we did not
attempt to optimize a data weighting scheme that fairly represented the
systematic uncertainties.  We did move to inclination angles that produced
just discernibly poorer fits to the velocity curves, then varied the central
mass to depart from the velocity amplitude by about 1 $\sigma$ at each
extremum. A more realistic disk model with finite
thickness would project more low radial velocity component into a given
pixel, suggesting that the mass estimated from a thin disk is systematically
too low.

How much of this nuclear mass can be attributed to luminous mass at the nucleus?
The WFPC2 continuum images in Paper~I show a bright nuclear point source
with V = $19\fm9$ (i.e., $L_V = 4.3 \times 10^6 \ L_{\odot}$ if 
$A_V$ to the nucleus is $0\fm54$; Paper~I) surrounded by the outer stellar distribution. Is the emission
from this continuum point source stellar or non-stellar? Additional 
spectroscopic observations of M84 are needed to address this issue. If we assume
that the emission arises entirely from a stellar population with
$M/L_V \approx 5$, the upper limit on the
luminous mass at the nucleus is $M_* \le 2 \times 10^7 \ M_{\odot}$.
This is much less than the nuclear mass $\approx 1.5 \times 10^9 \ M_{\odot}$
required to explain the gas disk dynamics (even given the uncertainty in
the extinction). Thus, M84 contains a dark compact mass (most
likely a BH given the presence of powerful radio jets).  
 
Although it was surprising that the emission-line isophotes did not describe
the geometry of the inner gas disk, the resulting picture is consistent
with physical expectations. The angular momentum axis of
the disk is now aligned with the axes of the dual radio jets. A gas disk
nearly perpendicular to the plane of the sky would suggest that these bipolar
radio jets are nearly in the plane of the sky, consistent with the roughly comparable power
in the jets and lobes on either side of the nucleus.  The more extended emission-line structure can now represent
a natural source of gas for an inner accretion region, without
the need for a strong, abrupt warp at $\sim 40$ pc from the center.
 
The BH mass estimate of $1.5 \times 10^9 \ M_{\odot}$ can be compared with the
expectation from Kormendy \& Richstone's (1995) correlation between black
hole mass and bulge mass.  An update to that relationship by
Kormendy et al.~(1997) including 12 BH mass determinations predicts
that the BH mass will be $0.0022^{+0.0014}_{-0.0009} \ \times$ the bulge mass.
For the total mass of M84, we use the structural parameter estimate of
Bender, Burstein, \& Faber (1992) of $M = 5G^{-1}\sigma_0^2r_e$.  We adopt
their value of effective radius (scaled to a distance of 17 Mpc) of 4.5 kpc,
and a central velocity dispersion of $\sim 310$ km~s$^{-1}$ from Davies \& Birkinshaw
(1988).  The resulting total galaxy mass is $5.1 \times 10^{11} \ M_{\odot}$, with a
predicted BH mass of $1.1 \times 10^9 \ M_{\odot}$ within a range of $(0.7-1.8) \times 10^9
\ M_{\odot}$.  Our best fit value of $1.5 \times 10^9 \ M_{\odot}$
is very close to the ridge line of the Kormendy
et al.~correlation.  At an $M_B = -21.0$, M84 is similar to M87 in total
absolute magnitude (therefore, total implied mass) and central BH
mass.

Our measurement of the BH mass in M84 (when combined with those for
M87 and NGC~4261; see \S 1) has interesting implications for the physics
of
radio galaxies. A comparison of these three measured BH masses with the total
radio luminosities of their associated radio sources (e.g., Roberts et al.~1991) suggests tentatively that these two quantities might not be directly
correlated. 
 
Several interpretive loose ends make M84 a galaxy meriting continuing study.
The shapes of the inner isophotes are not now explained straightforwardly
as a manifestation of an inner accretion disk.  What produces the change
in apparent P.A. and ellipticity?  Are there unresolved near-nuclear
H II regions that add to the complexity of the structure?  Is there any
relation between the sharp change in direction of the central dust lanes at $\sim 100$ pc
from the nucleus, as noted by Jaffe et al.~(1994) and in Paper~I, and the dynamical
environment produced by the central BH?  Although the best fitting systemic
velocity for the gas disk is consistent with the systemic velocity of
the outer filamentary gas derived from the two offset = $0\farcs0$ spectra, it
is $\sim 60$ km~s$^{-1}$ higher than the measured velocity for
the stars (e.g., Davies \& Birkinshaw 1988).  Stiavelli \& Setti (1993)
noted that galaxies with a high contrast central potential could produce
gravitational redshifts between central and outer samplings of the
stellar velocity field of several tens of km~s$^{-1}$ in extreme cases.
They found a $1.4\sigma$ effect in the Davies \&
Birkinshaw measurements of the stellar field.  If the dust lane obscuration
of the central stellar light is significant, it is possible
that the emission from the gas within 50 pc of the nucleus is coming from
much deeper in the central potential well, accounting for some of the
discrepancy.  Resolution of this velocity difference is necessary to add
confidence to the interpretation of the gas motion as that of a Keplerian
disk around a large unseen mass at the dynamical center of M84.

\acknowledgments

We acknowledge useful comments from Eric Emsellem and the referee Ralf Bender
and 
the assistance of P. Hall and C. Liu in planning
these observations.
Support for this work was provided to the STIS Investigation Definition
Team by NASA.

\clearpage

%
%

\par\noindent
{\bf References}

\par\noindent
Baum, S., et al.~1996, STIS Instrument Handbook, Version 1.0 
(Baltimore: STScI)
\par\noindent
Baum, S. A., et al.~1988, \apjs, 68, 643
\par\noindent
Baum, S. A., Heckman, T., \& van Breugel, W.
1990, \apjs, 74, 389
\par\noindent
Baum, S. A., Heckman, T., \& van Breugel, W.
1992, \apj, 389, 208
\par\noindent
Bender, R., Burstein, D., \& Faber, S. M.~1992, \apj, 399, 462
\par\noindent
Bower, G. A., Heckman, T. M., Wilson, A. S., \& Richstone, D. O.~1997, \apj,
483, L33 

(Paper~I)
\par\noindent
Burstein, D., \& Heiles, C. 1984,
\apjs, 54, 33
\par\noindent
Davies, R. L., \& Birkinshaw, M.~1988, \apjs, 68, 409
\par\noindent
Fanaroff, B. L., \& Riley, J. M.~1974, \mnras, 167, 31P
\par\noindent
Ferrarese, L., Ford, H. C., \& Jaffe, W.~1996, \apj, 470, 444 
\par\noindent
Hansen, L., N\o rgaard-Nielsen, H. U., \& 
J\o rgensen, H. E. 1985, \aap, 149, 442
\par\noindent
Harms, R. J., et al.~1994, \apj, 435, L35 
\par\noindent
Jaffe, W., et al.~1994, \aj, 108, 1567
\par\noindent
Jones, D. L., Sramek, R. A., \& Terzian, Y.~1981, ApJ, 246, 28
\par\noindent
Kormendy, J., Bender, R., Evans, A. S., \& Richstone, D.~1997, \aj, submitted
\par\noindent
Kormendy, J., \& Richstone, D.~1995, \araa, 33, 581
\par\noindent
Kormendy, J., et al.~1996, ApJ, 459, L57
\par\noindent
Laing, R. A., \& Bridle, A. H. 1987,
\mnras, 228, 557
\par\noindent
Mould, J., et al.~1995, ApJ, 449, 413
\par\noindent
Roberts, M. S., Hogg, D. E., Bregman, J. N., Forman, W. R., \& Jones, C.~1991,
\apjs, 75, 

751 
\par\noindent
Stiavelli, M., \& Setti, G. 1993, \mnras, 262, L51

%
%

\clearpage

\figcaption{Our STIS slit positions (with the solid lines representing the slit
edges) superposed on Paper~I's H$\alpha$ + [N~II]
image, which is displayed here with a logarithmic stretch with a
range in intensity covering a factor of 100. \label{fig1}}

\figcaption{A magnified view of the last offset $=0\farcs0$ spectrum
centered on the [N~II] $\lambda 6583$ emission line. To emphasize the
velocity gradient, the continuum distribution
has been subtracted, and the intensity scaling is logarithmic. The color
scale maps the range
of velocity along the slit, with
blue and red color representing velocities with respect to systemic that
are blueshifted and redshifted, respectively. The
dispersion axis (horizontal) covers a velocity interval of
1445 km~s$^{-1}$, while the spatial axis (vertical) covers the central
$3''$.
\label{fig2}}

\figcaption{Plots of the three spectral rows bracketing the nucleus
from the last spectrum obtained (offset = $0\farcs0$). The distance along the slit
$R$ is given in each panel, and the dashed lines represent the systemic
velocity of the [N~II] $\lambda 6583$ and [S~II] $\lambda 6731$ emission
lines. \label{fig3}}

\figcaption{Profiles of the [N II] $\lambda 6583$ emission line at positions
within $\sim 0\farcs3$ of M84's nucleus. The spectra have been continuum
subtracted and normalized to the peak intensity of [N~II] $\lambda 6583$. 
For each of the four slit positions,
the offset from the nucleus is given along the bottom, where negative
values of the offset move the slit toward P.A. = 14$\arcdeg$ on the sky. The position
along the slit (relative to the nucleus) in arcseconds is shown on the left,
where increasing values are toward P.A. = $104\arcdeg$. Each profile covers
a heliocentric velocity range of $200 - 2200$ km~s$^{-1}$, and the systemic
velocity is indicated by a dashed line. All profiles were taken from 
a single row in the data, except for the profiles at $+0\farcs30$ along
the slit which were binned by 4 pixels = $0\farcs2$ to improve the 
S/N. \label{fig4}}

\figcaption{The heliocentric velocity as a function of distance along the
slit for each slit position. Crosses indicate velocities measured by the
cross correlation technique described in the text. Closer to the nucleus
where two kinematic components are present, open triangles represent the
low-velocity component, and filled circles or squares represent the high-velocity
component. The squares denote positions where the high-velocity and low-velocity
components cannot be separated because both are at or near the systemic
velocity. The errors in the data points are $\le 25$ km~s$^{-1}$ (i.e., no
larger than the size of the points). The velocities predicted by the best
fit thin Keplerian disk model are represented by open circles connected by
a solid line.
\label{fig5}}

\clearpage

\begin{deluxetable}{lcc}
\footnotesize
\tablecaption{Keplerian Disk Model Parameters \label{tbl-1}}
\tablewidth{469.97pt}
\tablehead{
\colhead{Parameter} & \colhead{Best Fit}   & \colhead{Uncertainty Range}   
}
\startdata
Black Hole Mass ($M_{\odot}$) & $1.5 \times 10^9$ & $(0.9-2.6) \times 10^9$ \nl
Disk Inclination ($\arcdeg$) & 80 & $75-85$\tablenotemark{a} \nl
Disk P.A. ($\arcdeg$) & 83 & $80-85$ \nl
Gas systemic velocity (km~s$^{-1}$) & 1125 & $1100-1150$ \nl
Intensity law & $I(r) \propto r^{-1}$  \nl
$I(r)$ inner radius (pc) & 1 & $0.3-3$  \nl
$V(r)$ inner radius (pc) & 0.03 & $0.01-0.1$ \nl
PSF $\sigma \ ('')$ & $0.05$ & $0.04 - 0.06$ \nl

\enddata

\tablenotetext{a}{Lower mass requires lower inclination}

\end{deluxetable}

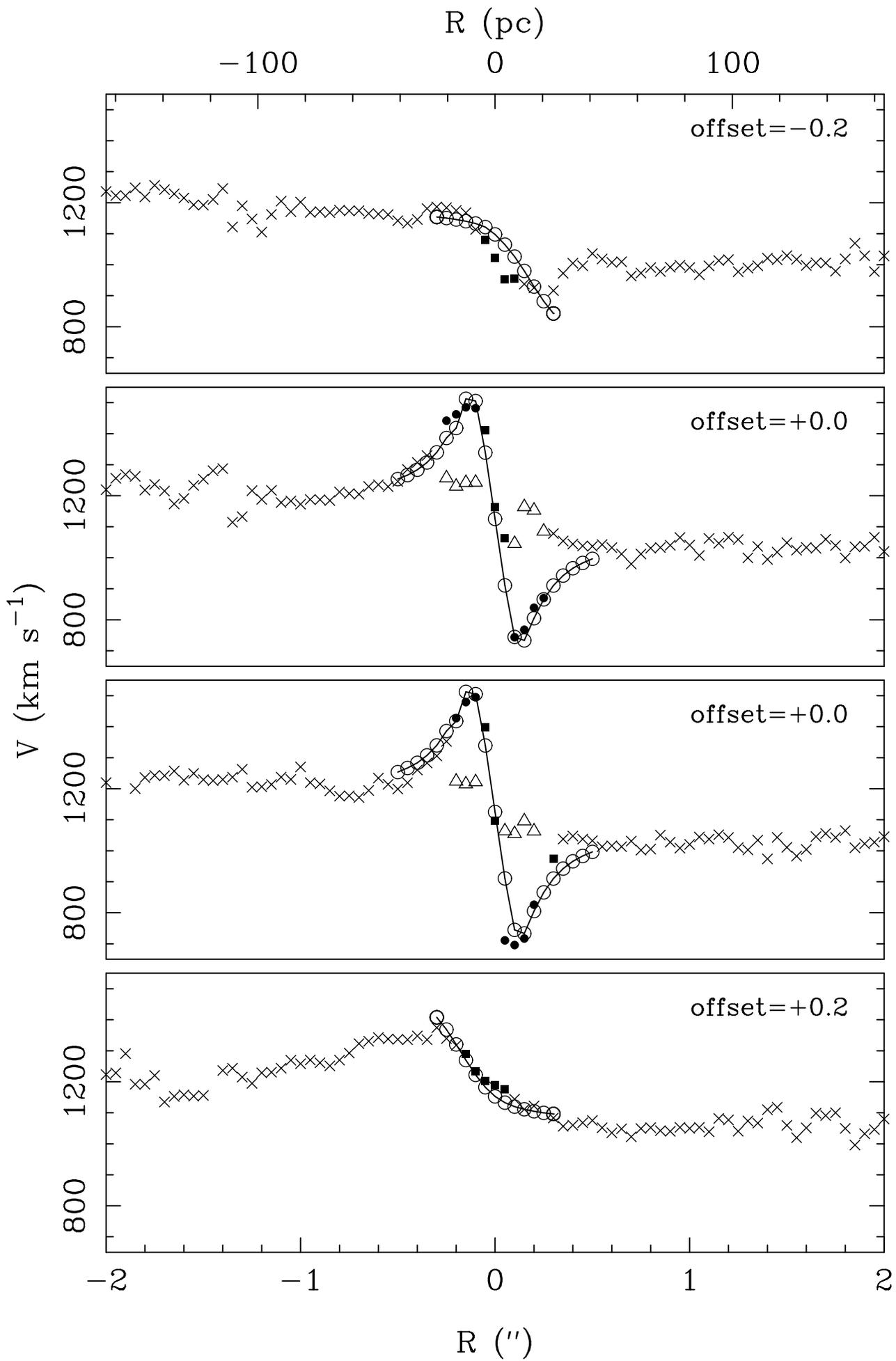
\end{document}